\documentclass[12pt]{article}
\usepackage{amsfonts}
\usepackage{euscript,amsmath,amssymb,epsfig,latexsym}
\usepackage[cp1251]{inputenc}
\usepackage[english]{babel}
\usepackage{float}
\newcommand{\be}{\begin{equation}}
\newcommand{\ee}{\end{equation}}
\newcommand{\bea}{\begin{eqnarray}}
\newcommand{\eea}{\end{eqnarray}}

\begin{document}

\title{\textbf{Cauchy Problem on \\
Non-globally Hyperbolic Spacetimes}}
\author{ I.Ya. Arefeva, T. Ishiwatari, I.V. Volovich \\
\, \\
\textit{Steklov Mathematical Institute} \\
\textit{Gubkin St. 8, 119991 Moscow, Russia} \\
\textit{arefeva@mi.ras.ru,takumi@mi.ras.ru}\\
\textit{volovich@mi.ras.ru}}
\date{~}
\maketitle

\begin{abstract}
Solutions of the Cauchy problem for the wave equation on a
 non-globally hyperbolic spacetime, which
contains closed timelike curves (time machines) are considered. It
is proved, that there exists a solution of the Cauchy problem, it is
discontinuous and in some sense unique for arbitrary initial
conditions, which are given on a hypersurface at time, that precedes
the formation of closed timelike curves (CTC). If the hypersurface of
initial conditions intersects the region containing CTC, then the solution of the Cauchy problem exists only for such
initial conditions, that satisfy a certain requirement of
self-consistency.
\end{abstract}

\newpage

\section{Introduction}

\label{s1 copy(1)}

There is a well developed theory of the Cauchy problem for hyperbolic
equations on globally hyperbolic spacetimes \cite{VS} - \cite{Leray}. A
spacetime oriented with respect to time (i.e. a pair $(M,g)$, where $M$ is a
smooth manifold, g is the Lorentz metric) is called globally hyperbolic, if $%
M$ is diffeomorphic to $\mathbb{R}^{1}\times \Sigma $, where $\Sigma
$ is a Cauchy surface. This definition is equivalent to the
definition of global hyperbolicity of Leray \cite{Leray,Hawking-Ellis}. Hyperbolic equations on non-globally hyperbolic
spacetimes have been considerably less studied, though numerous
examples of such spacetimes are described by such well known
solutions of equations for gravitation fields as solutions of
G\"{o}del, Kerr, Gott and many others \cite{Hawking-Ellis} -
\cite{Gott} (see \cite{Tod} for the history of constructing the G\"{o}del solution). Elementary examples of non-
globally hyperbolic spacetimes are the space $S_{t}^{1}\times \mathbb{R}
_{x}^{3}$ with Minkowski metric and the anti-de Sitter space. Let us note
that in those cases it is natural to examine the solutions of hyperbolic
equations with finite action \cite{KozVol}.

If a Lorentz manifold contains a closed timelike curve (time
machine), then it is not globally hyperbolic. There are several
papers in which simplest hyperbolic equations on non-globally
hyperbolic manifolds were discussed \cite{FMNEKTY} - \cite{Pol}. The
purpose of this work to study the wave equations on manifolds, which
contain closed timelike curves. We will consider a Minkowski plane
with two slits whose edges are glued in a specific manner. The
obtained manifold contains the conical points and the solution of
wave equation is, in general, discontinuous. We will prove that,
under natural conditions on jumps of right and left modes of the
solution of wave equation, the solution of Cauchy problem uniquely
exists, if given initial conditions on a line at a time, which
proceeds the formation of closed timelike curves (in this case, this
line is a Cauchy surface). If initial conditions are given on a
line, which intersects the region containing closed timelike curves
(in this case, this line is not a Cauchy surface), then the solution
of Cauchy problem exists only when a certain condition of
self-consistency is satisfied. In this case, additional initial
conditions are required for uniqueness of the solution of Cauchy
problem.

 Motivation of this work is related with the study of
possibility of creation of "wormholes" and mini time machines in
collisions of the particles at high energy \cite{TM}, see also
\cite{MMT}.

\newpage

\section{Discontinuous solutions of Cauchy problem.}

\label{s2 copy(1)}

In a halfplane $\mathbb{R}_{+}^{2}=\{(t,x)\in \mathbb{R}^{2}|t>0\},$ let us
consider two vertical intervals $\gamma _{1}$ and $\gamma _{2}$ with length $%
l>0$:

\begin{equation}
\begin{aligned} &\gamma_1=\{(t,x)\in\mathbb R_+^2\mid x=a_1,
\,b_1<t<b_1+l\}, \\[2mm] &\gamma_2=\{(t,x)\in\mathbb R_+^2\mid x=a_2,
\,b_2<t<b_2+l\}. \end{aligned}  \label{1}
\end{equation}
We assume,
\begin{equation*}
a_{2}>a_{1},\qquad b_{2}>b_{1}+l+a_{2}-a_{1}.
\end{equation*}

Suppose, that the edges of the two slits are glued as illustrated in
Fig.1~\cite{Deut},~\cite{Pol}. The obtained regions has two singular points -
conical singularity - the ends of slits. 

\begin{figure}[]
$\,\,\,\,\,\,\,$$\,\,\,\,\,\,\,$$\,\,\,\,\,\,\,$ $\,\,\,\,\,\,\,$$%
\,\,\,\,\,\,\,$$\,\,\,\,\,\,\,$ $\,\,\,\,\,\,\,$$\,\,\,\,\,\,\,$$%
\,\,\,\,\,\,\,$
\includegraphics[height=2.5in]{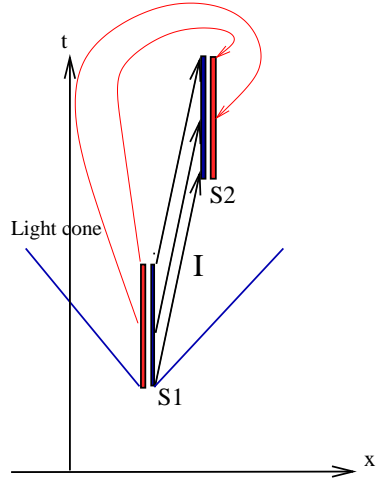}
\caption{{\small Minkowski plane with two segments, being glued in a
specific manner : "inner" edges of the two slits are
glued to each other, "outer" edges of the two slits  are
also glued to each other. The points of identification on "outer"
edges of the two slits are indicated by  lines with an arrow, the
points of identification on "inner" edges of the two slits are
indicated by  lines with an arrow. On the picture, the light
cone is drawn by  lines, going out from the point $S_{1}$ with
the coordinate $(a_{1,}b_{1})$. It is assumed, that the vector $I$,
which realizes the identification, is timelike and it forms a closed timelike curve. The point $S_{2}$
has the coordinate $(a_{2},b_{2})$. }} \label{slits1}
\end{figure}

Let us consider Cauchy problem for the wave equation in an open subset (see
below) of the region $\mathbb{R}_{+}^{2}\backslash \{\overline{\gamma }%
_{1}\cup \overline{\gamma }_{2}\}$ for functions $u=u(t,x)$%
\begin{align}
& u_{tt}-u_{xx}=0,  \label{2} \\[2mm]
& u(t,x)|_{t=0}=u_{0}(x),\qquad \partial _{t}u(t,x)|_{t=0}=u_{1}(x),
\label{3}
\end{align}%
Here $\overline{\gamma }_{i}$ -- are closures of the  intervals $\gamma _{i}$, $i=1,2.$

Suppose, function $u(t,x)$ and its first derivative with respect to $t$ and $%
x$ admit a continuous extension on the intervals $\gamma _{1}$ and $\gamma
_{2}$ when $(t,x)$ goes to $\gamma _{1}$ and $\gamma _{2}$ from right and
left, and we set the following identification  conditions 

\begin{align}
& u(t,x)|_{x=a_{1}-0}=u(t+b_{2}-b_{1},x)|_{x=a_{2}+0},  \label{4} \\[1mm]
& \partial _{x}u(t,x)|_{x=a_{1}-0}=\partial
_{x}u(t+b_{2}-b_{1},x)|_{x=a_{2}+0},  \label{5} \\[1mm]
& u(t,x)|_{x=a_{1}+0}=u(t+b_{2}-b_{1},x)|_{x=a_{2}-0},  \label{6} \\[1mm]
& \partial _{x}u(t,x)|_{x=a_{1}+0}=\partial
_{x}u(t+b_{2}-b_{1},x)|_{x=a_{2}-0},  \label{7}
\end{align}%
where $b_{1}<t<b_{1}+l$.

As is well-known \cite{VS}, the change of variables such that

\begin{equation}
\tilde{u}(\xi ,\eta )=u\biggl(\frac{\eta -\xi }{2},\frac{\eta +\xi }{2}%
\biggr),\qquad \xi =x-t,\quad \eta =x+t,  \label{8}
\end{equation}%
transforms the equation \eqref{2} to the canonical form
\begin{equation}
\frac{\partial ^{2}\tilde{u}(\xi ,\eta )}{\partial \xi \partial \eta }=0,
\label{WE}
\end{equation}%
from which it follows that the solution of the equation (\ref{2}) is
given by

\begin{equation*}
u(t,x)=f(x-t)+g(x+t),
\end{equation*}%
where the functions $f(\xi )$ and $g(\eta )$ belong to the class $C^{2}$ on
the corresponding intervals for variables $\xi $ and $\eta $.

Let us consider\ the following problem. Suppose disconnected region $\Omega
_{-}\subset \mathbb{R}_{+}^{2}$ is given by
\begin{equation}
\Omega _{-}=\mathcal{D}_{1}\cup \mathcal{D}_{2}\cup \mathcal{D}_{0},
\label{10}
\end{equation}%
where the regions $\mathcal{D}_{1}$, $\mathcal{D}_{2}$ and
$\mathcal{D}_{0}$ are bounded by closed intervals $\gamma _{1}$,
$\gamma _{2}$ and half-lines (rays), going out from the ends of
intervals $\gamma _{1}$, $\gamma _{2}$ to the right by angle
$45^{o}~$ (see Fig.2).

\begin{figure}[]$\,\,\,\,\,\,\,\,\,\,\,\,\,\,\,\,\,\,\,\,\,\,\,\,\,\,$
\begin{picture}(70,200)
\put(0,0){\includegraphics[height=2.5in]{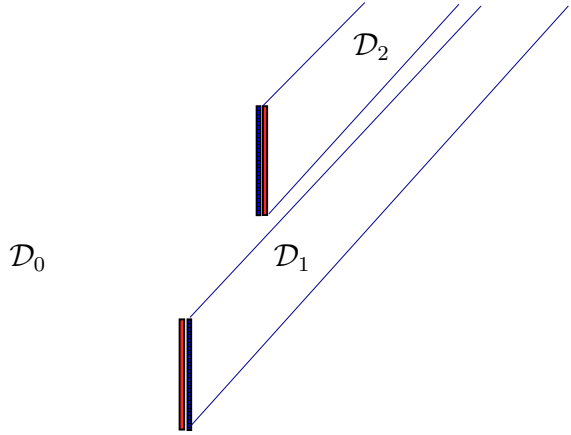}}
\put(180,80){${\cal D}_1$} \put(210,160){${\cal D}_2$}
\put(80,80){${\cal D}_0$}
\end{picture}
\caption{{\small Region} $\Omega _{-}=\mathcal{D}_{1}\cup \mathcal{D}_{2}\cup
\mathcal{D}_{0}$} \label{cauchi-R}
\end{figure}

Any function $w$ of class $C^{k}(\Omega _{-})$ consists of three components $
w=(w_{0},w_{1},w_{2})$, where $w_{i}\in C^{k}(\mathcal{D}_{i}),~i=0,1,2.$

Let us recall, that $C^{k}(\overline{\mathcal{D}})$ means the class of
functions $C^{k}(\mathcal{D})$, which together with its first derivative
admit a continuous extension on the closure $\overline{\mathcal{D}}$
\cite{VS}.

Let us denote by $\mathcal{F}^{1}(\Omega _{-})$ the class of such functions $%
w=(w_{0},w_{1},w_{2})$ belonging to $C^{1}(\Omega _{-})$, that $w_{i}\in
C^{1}(\overline{\mathcal{D}}_{i}),i=0,1,2.$\bigskip

\textbf{Problem \ }\ \emph{Find a function }$u(t,x)\in \mathcal{F}^{2}(\Omega _{-})$%
\emph{, where }$\Omega _{-}$\emph{\ is a disconnected open set given by
\eqref{10}, which satisfies the equation}
\begin{equation}
\frac{\partial \tilde{u}(\xi ,\eta )}{\partial \eta }=0,\qquad (\xi ,\eta
)\in \Omega _{-}  \label{11}
\end{equation}%
\emph{(and, therefore, satisfies the wave equation \eqref{WE}), also
satisfies the identification  conditions  \eqref{4} - \eqref{7}
and
the initial condition such that}\textit{\ }%
\begin{equation}
u(t,x)|_{t=0}=u_{0}(x),\qquad x\in \mathbb{R}.  \label{12}
\end{equation}

\bigskip

\textbf{Proposition\  }\label{pr1}

\emph{Let} \textit{$u_{0}\in C^{2}(\mathbb{R})$,}\emph{\ then the solution
of the problem above, belonging to the class }$\mathcal{F}^{2}(\Omega _{-}),$\emph{\
exists~and it is unique}.

\textbf{Proof.} From the equation \eqref{11}\ it follows, that in
any one of regions $\mathcal{D}_{i}$, $i=0,1,2,$ the solution is
given by the formula
\begin{equation}
u(t,x)=f_{i}(x-t),\qquad (t,x)\in \mathcal{D}_{i},\quad i=0,1,2,  \label{13}
\end{equation}%
where $f_{i}(\xi )$ is a function belonging to the class $C^{2}$ on the
region, where the argument $\xi ~$varies.~From the initial condition
\eqref{12} it follows%
\begin{equation}
f_{0}(x)=u_{0}(x),\qquad x\in \mathbb{R}.  \label{14}
\end{equation}

\begin{figure}[h]$\,\,\,\,\,\,\,\,\,\,\,\,\,\,\,\,\,\,\,\,\,\,\,\,\,\,$\begin{picture}(70,200)
\put(0,0){\includegraphics[height=2.5in]{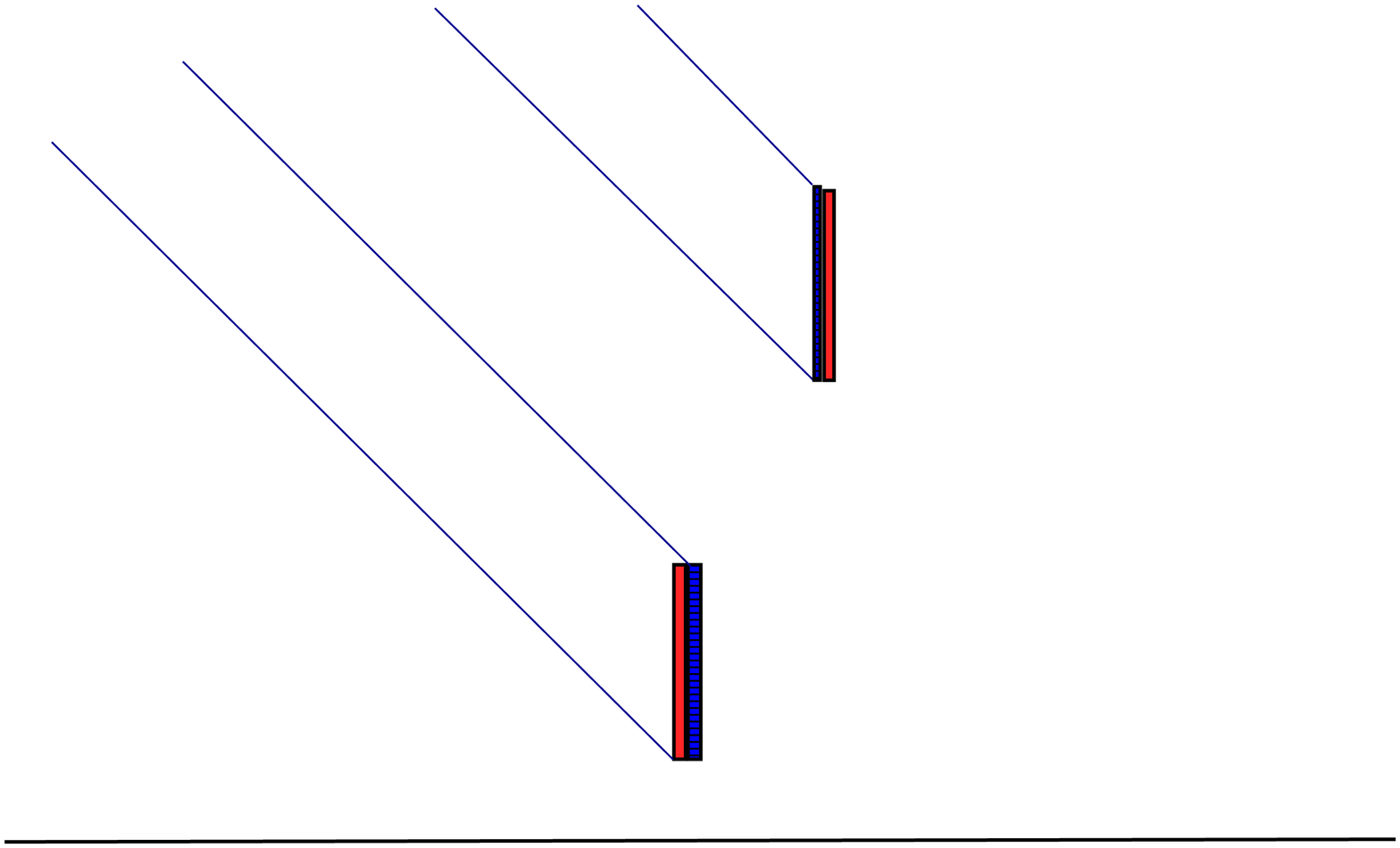}}
\put(90,80){${\cal D}_3$} \put(120,160){${\cal D}_4$}
\put(200,80){${\cal D}^\prime_0$}
\end{picture}
\caption{$\Omega_+={\cal D}_3\cup{\cal D}_4\cup{\cal D}^\prime_{0}$}
\label{fig2}
\end{figure}

Further, by using the  identification condition \eqref{4} , we
have
\begin{equation}
u_{0}(a_{1}-t)=f_{2}(a_{2}-b_{2}+b_{1}-t),\qquad b_{1}<t<b_{1}+l,  \label{15}
\end{equation}%
while the identification  condition \eqref{6} gives
\begin{equation}
f_{1}(a_{1}-t)=u_{0}(a_{2}-b_{2}+b_{1}-t),\,\,\,b_{1}<t<b_{1}+l.  \label{s1}
\end{equation}%
Let us note that the identification  conditions \eqref{5} and
\eqref{7}
 for functions in the form \eqref{13} are obtained from the conditions
\eqref{4} and \eqref{6}, respectively.

\ \ If $(t,x)\in D_{1}$, then $\xi =x-t~$ varies in the interval
(Fig.3) such that
\begin{equation}
a_{1}-b_{1}-l<\xi <a_{1}-b_{1},  \label{17}
\end{equation}%
on which the function $f_{1}(\xi )$ is defined.

\ \ If $(t,x)\in D_{2}$, then $\xi =x-t~$ varies in the interval such that
\begin{equation}
a_{2}-b_{2}-l<\xi <a_{2}-b_{2},  \label{18}
\end{equation}%
on which the function$f_{2}(\xi )$~is defined.

\ \ If $(t,x)\in D_{0}$, then $\xi =x-t~$ varies on the whole real axis $%
\mathbb{R}$ (Fig.\ref{fig2}).

Formula (\ref{s1}) gives a function such that
\begin{equation}
f_{1}(\xi )=u_{0}(\xi +a_{2}-b_{2}+b_{1}-a_{1}),  \label{19}
\end{equation}%
where $\xi =x-t$ varies on the interval \eqref{17}.

Likewise, formula \eqref{15} defines a function such that%
\begin{equation}
f_{2}(\xi )=u_{0}(\xi +a_{1}-a_{2}+b_{2}-b_{1}),  \label{20}
\end{equation}%
where $\xi =x-t$ varies on the interval \eqref{18}.

Thus, the solution of the problem  is given by the formulas
\eqref{13}, where the functions $f_{0},~$ $f_{1}~$and$~~f_{2}$ are
given by \eqref{14}, \eqref{19} and \eqref{20}.

Proposition is proved.\bigskip

\ \ The following statement is derived\ from the proof of the Preposition.

\textbf{Statement 1 \ } \textit{If the function} \textit{\
$u=u(t,x)\in \mathcal{F}^{1}(\Omega _{-})$ satisfies the equation
\eqref{11} and the identification conditions  \eqref{4} -
\eqref{7}, then
$u(x,t)$ is given by the formulas }%
\begin{equation}
\begin{alignedat}{2} u(t,x)&=v(x-t), &\qquad (t,x)&\in\mathcal D_0, \\[-1mm]
u(t,x)&=v(x-t+a_2-a_1-b_2+b_1),&\qquad (t,x)&\in\mathcal D_1, \\[-1mm]
u(t,x)&=v(x-t+a_1-a_2+b_2-b_1),&\qquad (t,x)&\in\mathcal D_2, \end{alignedat}
\label{22}
\end{equation}%
\textit{where $v$ is some function belonging to the class} $C^{1}(\mathbb{R})
$.\bigskip

\textbf{Remark 1} \ \ Let us note, that some sums of jumps on the
discontinuities are equal to zero. For example, $\Delta
_{S_{1}}+\Delta
_{S_{2}}=0,~$where $\Delta _{S_{i}}=u(t+0,x)-u(t-0,x)~$for~$%
t-x=b_{i}-a_{i},~i=1,2$ (see Fig. \ref{cauchi-R}).

\bigskip

A similar statement holds for $\Omega _{+}=\mathcal{D}_{3}\cup \mathcal{D}
_{4}\cup \mathcal{D}_{0}^{\prime }$, see Fig. 3.

\textbf{Statement 2} \ \ \ \textit{If the function\ $u=u(t,x)\in \mathcal{F}
^{1}(\Omega _{+})$, where $\Omega _{+}=\mathcal{D}_{3}\cup \mathcal{D}
_{4}\cup \mathcal{D}_{0}^{\prime }$ - disconnected open set shown at Fig.3,
satisfies the equation }
\begin{equation}
\frac{\partial \tilde{u}(\xi ,\eta )}{\partial \xi }=0,\qquad (\xi ,\eta
)\in \Omega _{+},  \label{23}
\end{equation}
\textit{and the conditions for identification \eqref{4} -
\eqref{7}, then $u(x,t)$ is given by the formulas }
\begin{equation}
\begin{alignedat}{2} u(t,x)&=v(x+t),&\qquad (t,x)&\in\mathcal D'_0, \\
u(t,x)&=v(x+t+a_2-a_1+b_2-b_1),&\qquad (t,x)&\in\mathcal D_3, \\[-1mm]
u(t,x)&=v(x+t+a_1-a_2+b_1-b_2),&\qquad (t,x)&\in\mathcal D_4, \end{alignedat}
\label{24}
\end{equation}
\textit{where $v$ is some function belonging to the class} $C^{1}(\mathbb{R})
$.\bigskip

The following theorem is valid.

\textbf{Theorem 1.\thinspace } \textit{Given functions $u_{0}\in C^{2}(
\mathbb{R})$ and $u_{1}\in C^{1}(\mathbb{R})$. Let functions $f$ and $g$ be
defined by the following formulas}

\begin{equation}
f(x)=\frac{1}{2}\biggl[u_{0}(x)-\int_{x_{0}}^{x}u_{1}(s)\,ds\biggr],\qquad
g(x)=\frac{1}{2}\biggl[u_{0}(x)+\int_{x_{0}}^{x}u_{1}(s)\,ds\biggr],
\label{25}
\end{equation}
\textit{where $x_{0}\in \mathbb{R}$. \ Let us divide the region $\mathbb{R}
_{+}^{2}\backslash \{{\overline{\gamma }}_{1}\cup \overline{\gamma }_{2}\}$
into regions $\mathcal{D}_{i}$, $i=1,...7$ as shown in Fig. 4. Then
the components of the functions $u(t,x)$, which are given in the regions $
\mathcal{D}_{i}$ by the following formulas}
\begin{equation}
\begin{alignedat}{2} u(t,x)&=f(x-t+a_2-a_1+b_1-b_2)+g(t+x), &\qquad
(t,x)&\in\mathcal D_1, \\ u(t,x)&=f(x-t+a_1-a_2+b_2-b_1)+g(t+x), &\qquad
(t,x)&\in\mathcal D_2, \\ u(t,x)&=f(x-t)+g(t+x+a_2-a_1+b_2-b_1), &\qquad
(t,x)&\in\mathcal D_3, \\ u(t,x)&=f(x-t)+g(t+x+a_1-a_2+b_1-b_2), &\qquad
(t,x)&\in\mathcal D_4, \\ u(t,x)&=f(x-t)+g(t+x), &\qquad (t,x)&\in\mathcal
D_i, \quad i=5,6,7, \end{alignedat}  \label{26}
\end{equation}
\textit{belong to the class} $C^{2}(\mathcal{D}_{i})\cap C^{1}(\overline{
\mathcal{D}}_{i})$, $i=1,2,\dots ,7~$\textit{and they} \textit{are the
solutions of the wave equation \eqref{2} in the regions} \textit{$\mathcal{
D}_{i}$ and satisfy the initial condition \eqref{3} and the
identification conditions   \eqref{4} - \eqref{7}}.

\begin{figure}[]$\,\,\,\,\,\,\,\,\,\,\,\,\,\,\,\,\,\,\,\,\,\,\,\,\,\,$
\begin{picture}(70,200)
\put(0,0){\includegraphics[height=2.5in]{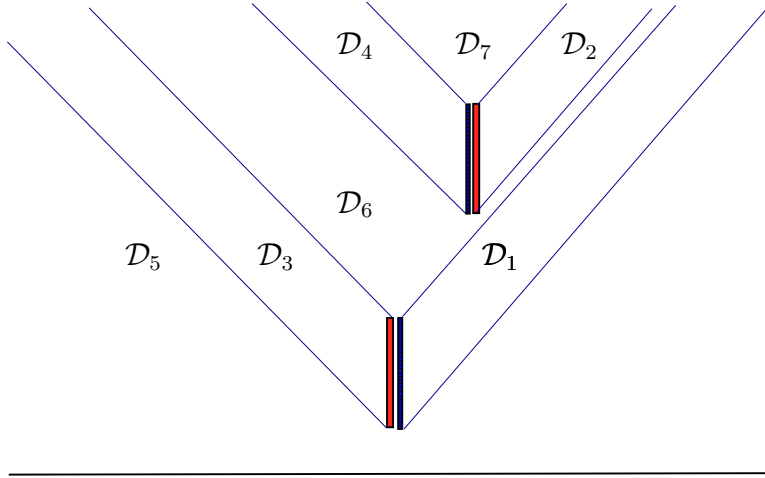}}
\put(125,100){${\cal D}_6$}\put(125,160){${\cal D}_4$}
\put(210,160){${\cal D}_2$}\put(170,160){${\cal D}_7$}
\put(180,80){${\cal D}_1$}\put(180,80){${\cal D}_1$}
\put(95,80){${\cal D}_3$} \put(45,80){${\cal D}_5$}
\end{picture}
 \caption{{\small Regions}
$\mathcal{D}_{i}$, i=1,2,...7} \label{cauchi-CTC-simple}
\end{figure}

\bigskip

\textit{Proof }of the theorem is straightforward.

\bigskip

\section{ Minimally discontinuous solutions of wave \\ equation}

Consider now the problem on uniqueness of the solution of Cauchy problem for
the wave equation on disconnected regions.

Suppose, for each region $D_{i},~i=1,...,7,\ $which is introduced in the
Fig. 4, the solution of the wave equation belong to the class $C^{2}(D_{i}).~$
Then, by the well known theorem \cite{VS}, in each $D_{i}~$the
solution  is given by a sum of the right mode and the left mode.

We say, that a function, which satisfies the wave equation in disconnected
regions $D_{i},~$satisfies the condition of "minimal discontinuity", if the
right and left mode, which are defined in $D_{i},$ admit the extension onto
the region $\mathcal{D}_{0}~$and $\mathcal{D}_{0}^{\prime },~$respectively$.$

For solutions \eqref{26}, let us define functions $u_{r}(t,x)$ (right
mode);

\begin{equation}
\begin{alignedat}{2} &u_\mathrm{r}(t,x)=f(x-t+a_2-a_1+b_1-b_2), &\qquad
(t,x)&\in \mathcal D_1, \\ &u_\mathrm{r}(t,x)=f(x-t+a_1-a_2+b_2-b_1),
&\qquad (t,x)&\in\mathcal D_2, \\ &u_\mathrm{r}(t,x)=f(x-t), &\qquad
(t,x)&\in\mathcal D_i, \quad i=3,4,5,6,7, \end{alignedat}  \label{27}
\end{equation}

and functions $u_{l}(t,x)$ (left mode)

\begin{equation}
\begin{alignedat}{2} &u_\mathrm{l}(t,x)=g(t+x),&\qquad (t,x)&\in\mathcal
D_i,\quad i=1,2,5,6,7, \\ &u_\mathrm{l}(t,x)=g(t+x+a_2-a_1+b_2-b_1), &\qquad
(t,x)&\in\mathcal D_3, \\ &u_\mathrm{l}(t,x)=g(t+x+a_1-a_2+b_1-b_2), &\qquad
(t,x)&\in\mathcal D_4. \end{alignedat}  \label{28}
\end{equation}

Then, the function $u_{r}(t,x)$ (right mode), which is defined on
the region $\mathcal{D}_{i}$ by the formulas \eqref{27}, admits a
continuous extension onto the region $\mathcal{D}_{0}$ (see
Fig.\ref{fig2}), while the
function $u_{l}(t,x)$ (left mode), which are defined in the region $\mathcal{
D}_{i}$ by the formulas \eqref{28}, admits a continuous extension
onto the region $\mathcal{D}_{0}^{\prime }$ (see
Fig. 3).\bigskip

\textbf{Theorem 2} \ \ \ \textit{With the assumption of "minimal
discontinuity", the solution of the problem 
\eqref{2} - \eqref{7} is unique and is given by the formulas
\eqref{26}.}

\textbf{Remark 2} \ \ \ It is possible to say, that one can consider the theorem 2 as an analogy to the~edge~of~the~wedge theorem in the
theory of functions of complex variables \cite{VS-CV}.

\textbf{Remark 3} \ \ \ It may be interesting to note that when the
two slits are at the same height (Fig. 5), $b_{1}=b_{2}~$ and
none of them is in the shadow of the other one, then the solutions
corresponding to the problem \eqref{2}, \eqref{3}, \eqref{4} -
\eqref{7} are given by the formulas, which are similar to \eqref{26}
with $b_{1}=b_{2}$.

\begin{figure}[]$\,\,\,\,\,\,\,\,\,\,\,\,\,\,$
\includegraphics[height=2in]{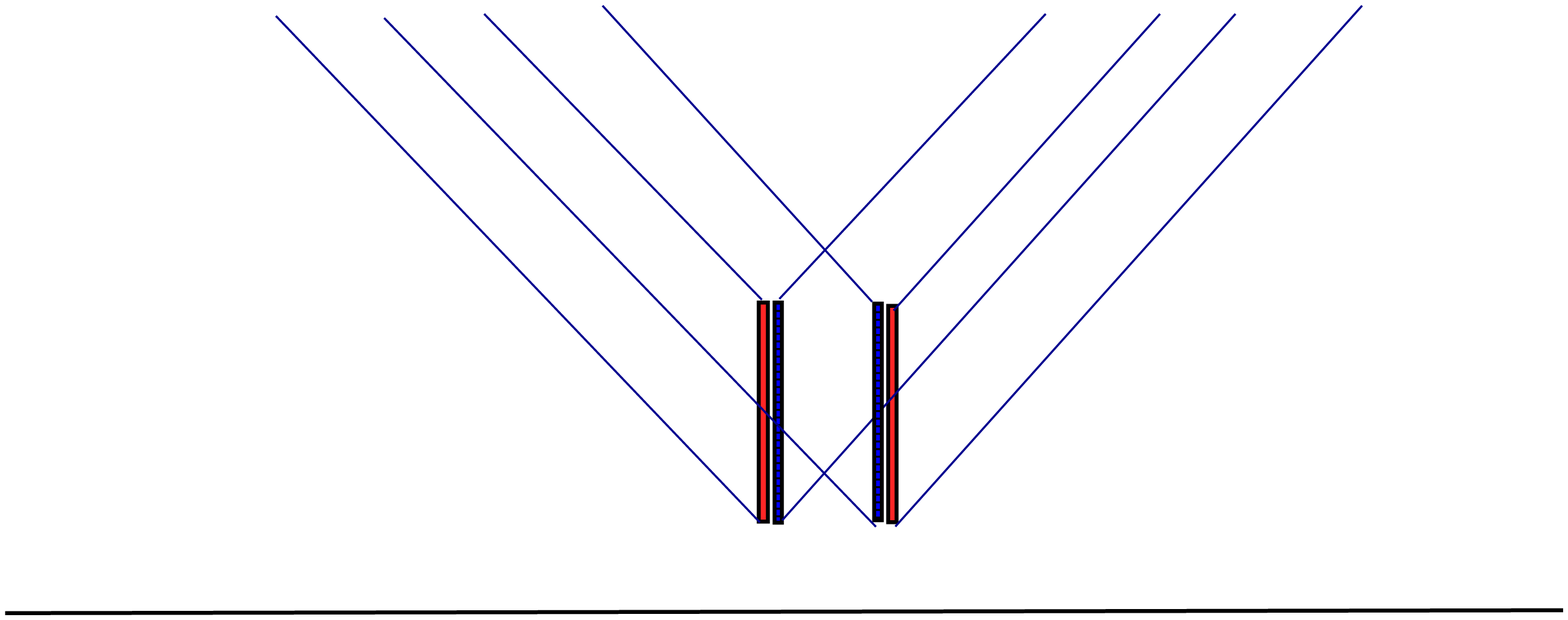}
\caption{{\small Two slits are at the same height and none of them is in the
shadow of the other one. The vector of identification is spacelike.
}}
\label{cauchi0-m 1}
\end{figure}

The condition of "minimal discontinuity" can be formulated in the terms of
the original function $u(t,x)$. We say, a solution satisfies the requirement
of the condition of "minimal discontinuity" with an accuracy of jumps by
constant", if the solution has left continuous derivatives (derivatives with
respect to $x+t$) on the right-hand side of the light cones, the tops of
which are located at the point $S_{i}$ and $T_{i}$, $i=1,2$,
\begin{equation}
\partial _{t+x}u(t,x)|_{x-t=a_{i}-b_{i}+0}=\partial
_{t+x}u(t,x)|_{x-t=a_{i}-b_{i}-0},\qquad t>x,  \label{29}
\end{equation}
and, at the same time, the solution has also left continuous derivatives
(derivatives with respect to $x-t$) on the left-hand side of the light
cones, the tops of which are located at the point $S_{i}$ and $T_{i}$, $i=1,2
$,
\begin{equation}
\partial _{x-t}u(t,x)|_{x+t=a_{i}+b_{i}+0}=\partial
_{x-t}u(t,x)|_{x+t=a_{i}+b_{i}-0},\qquad t>x.  \label{30}
\end{equation}
If we assume that the right mode and the left mode decrease, then
from the conditions \eqref{29} and \eqref{30} follows  the
minimal discontinuous condition.\bigskip

\section{Selfconsistent initial values of Cauchy problem}

It is important to note, that for the spacetime shown in Fig. 1, it
is not possible to formulate the Cauchy problem for any variable $t=t_{0}.$
Let us recall, that the Cauchy surface in the spacetime $(M,g)~$is such a
hypersurface $\Sigma ,$ that every geodetic,$~$which is released from any
point of $M,~$intersects the hypersurface$~\Sigma $ once and only once.

In particular, for the line $C_{b}~$illustrated in Fig. 6, we can't
give the right modes on the segment $c$ independently of those on the
segment $c^{\prime }$ and we can't give the left modes on the segment $d$
independently of those on the segment $d^{\prime }$.

\begin{figure}[tbp]
\begin{picture}(70,200)
\put(0,0){\includegraphics[height=3.5in]{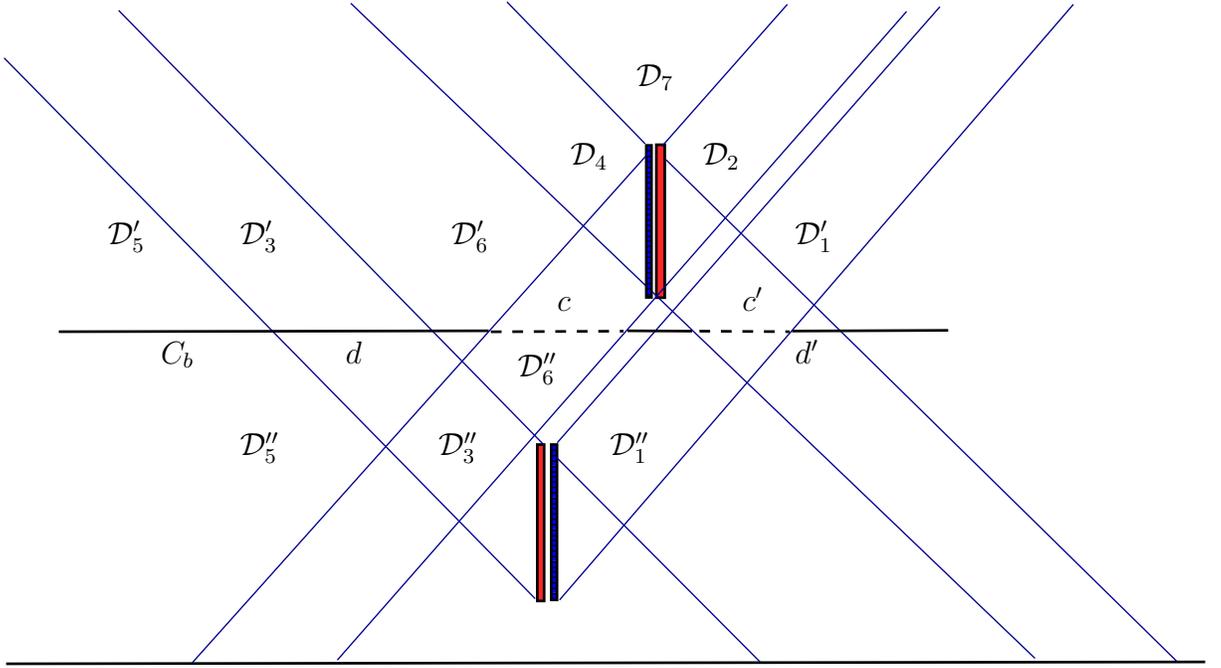}}
\put(240,220){${\cal D}_7$}
\put(215,190){${\cal D}_4$}
\put(265,190){${\cal D}_2$}
\put(40,160){${\cal D}^\prime_5$}\put(90,160){${\cal D}^\prime_3$}
\put(170,160){${\cal D}^\prime_6$}\put(300,160){${\cal D}^\prime_1$}
\put(210,135){$c$}\put(280,135){$c^\prime$}
\put(195,110){${\cal D}^{\prime\prime}_6$}
\put(60,115){$C_b$}\put(130,115){$d$}\put(300,115){$d^\prime$}
\put(230,80){${\cal D}^{\prime\prime}_1$}\put(90,80){${\cal D}^{\prime\prime}_5$}
\put(165,80){${\cal D}^{\prime\prime}_3$}
\end{picture}
\caption{{\small The surface $C_{b}$ divides the region $\mathcal{D}_{i}$,
$i=1,3,5,6
$, illustrated in Fig. \protect\ref{cauchi-CTC-simple}, into the regions $
\mathcal{D}_{i}^{\prime }$ and $\mathcal{D}_{i}^{\prime \prime }$, $i=1,3,5,6
$, respectively.  It is not possible to formulate the
problem (\protect\ref{2}), (\protect\ref{3}) with the identification conditions
(\protect\ref{4}) - (\protect\ref{7}) and with
arbitrary initial conditions on the surface $C_{b}$.}}
\label{cauchi-CTC-Dr-Dl}
\end{figure}

This is related to the fact, that the right modes, which satisfy
equation \eqref{11} and also the conditions for identification
\eqref{4} - \eqref{7}, do not exist for arbitrary initial
conditions. The conditions for identification \eqref{4},   \eqref{6}
require that the initial conditions on the segments $c$ and
$c^{\prime }$ should coincide
with each other, see Fig. 7. However, these initial conditions on  
$C_{b}$ don't define the solution on the region $\mathcal{D}_{2}$.

A similar property appears also for the wave equation with the initial
conditions on the surface $C_{b}$ as shown in Fig. 6.

\begin{figure}[]\begin{picture}(70,200)
\put(0,0){\includegraphics[height=3.5in]{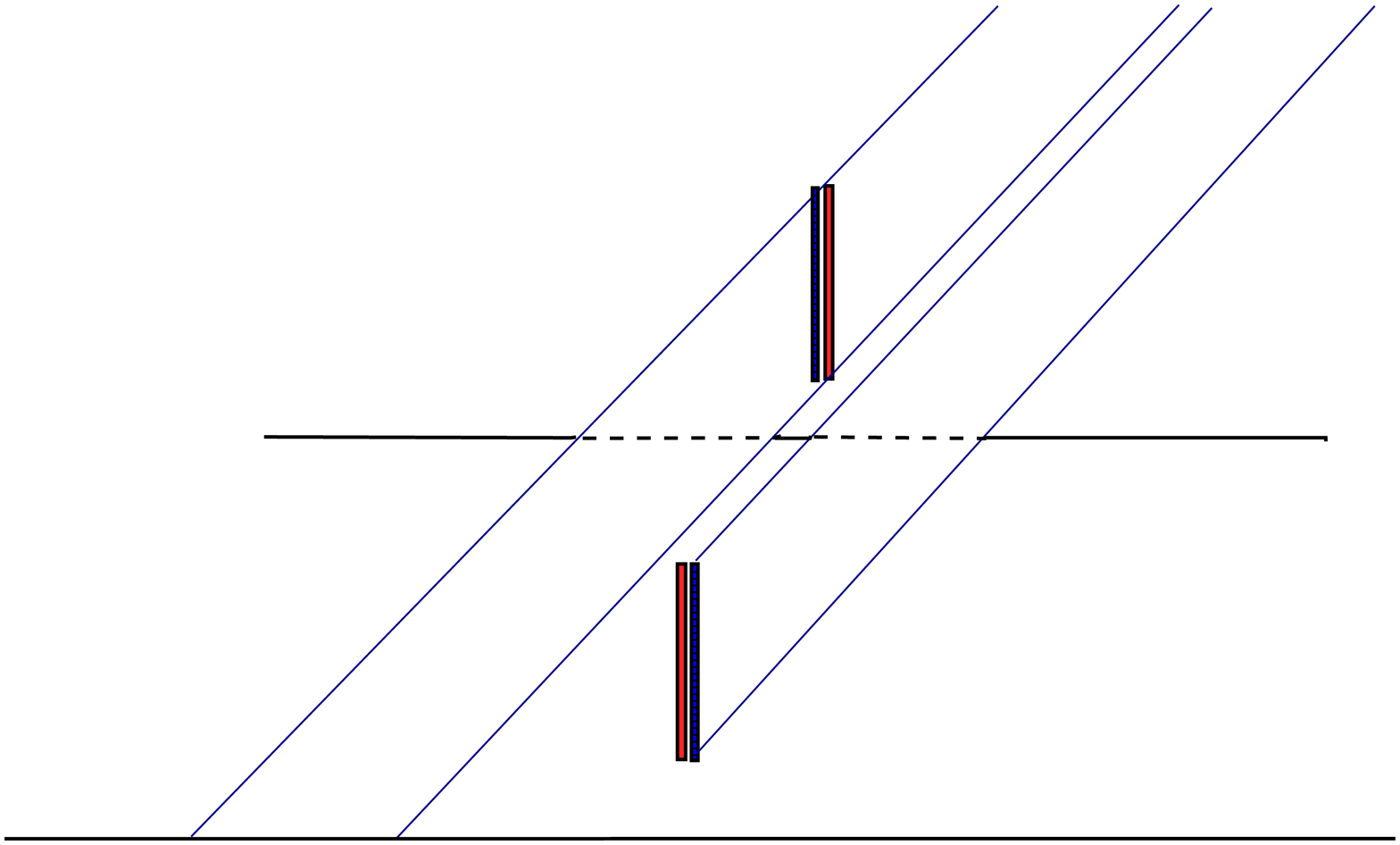}}
\put(265,190){${\cal D}_2$}
\put(40,120){$C_b$}
\put(170,160){${\cal D}^\prime_0$}\put(300,160){${\cal D}^\prime_1$}
\put(210,135){$c$}\put(280,135){$c^\prime$}
\put(230,80){${\cal D}^{\prime\prime}_1$}\put(90,80){${\cal D}^{\prime\prime}_0$}
\end{picture}
\caption{{\small The surface $C_{b}$ divides the region $\mathcal{D}_{i}$,
$i=0,1$,
illustrated in Fig.\protect\ref{cauchi-R}, into the regions $\mathcal{D}
_{i}^{\prime }$ and $\mathcal{D}_{i}^{\prime \prime }$, $i=0,1$,
respectively.  It is not possible to formulate the
problem (\protect\ref{11}) with the identification conditions  (
\protect\ref{4}), (\protect\ref{6}) and with arbitrary initial
conditions of the form (\protect\ref{12}) on the surface $C_{b}$.}}
\label{cauchi-CTC-Dr-Dl-r}
\end{figure}

\newpage
\section*{Acknowledgements}
                                
This paper is dedicated to V.S.Vladimirov with best wishes on the
85th birthday.
 I.A. is
partly supported by RFFI grant 08-01-00798 and grant NS-795.2008.1.
 I.V.
is partly supported by RFFI grant 08-01-00727,  NSh-3224.2008.1, DFG Project 436 RUS 113/951. T.I. is partially supported by RFFI grant 08-01-00727.  I.A. would like
to thank H.Nielson for fruitful discussions.

\end{document}